\documentclass[11pt]{article}
\usepackage{amsfonts}
\usepackage{graphicx}
\usepackage{epstopdf}
\usepackage{epsfig}
\usepackage{amssymb}
\usepackage{setspace}
\usepackage{caption}
\usepackage{amsfonts}
\usepackage{color}
\usepackage{amsmath}
\usepackage{float}
\usepackage{comment}
\usepackage{subcaption}
\newcommand {\be}{\begin{equation}}
\newcommand {\ee}{\end{equation}}
\newcommand {\bea}{\begin{array}}
	
	\newcommand {\eea}{\end{array}}

\textwidth=6.5in \hoffset=-.75in \voffset=-1in \numberwithin{equation}{section}
\numberwithin{figure}{section}

\begin{document}

	\begin{titlepage}
		\vspace{1cm}
		\begin{center}
			{\Large \bf {Melvin-Zipoy-Voorhees Spacetime and Circular Orbits}}\\
		\end{center}
		\vspace{2cm}
		\begin{center}
			\renewcommand{\thefootnote}{\fnsymbol{footnote}}
			Haryanto M. Siahaan{\footnote{haryanto.siahaan@unpar.ac.id}}\\Program Studi Fisika, Universitas Katolik Parahyangan\\
			Jalan Ciumbuleuit 94, Bandung 40141, Indonesia
			\renewcommand{\thefootnote}{\arabic{footnote}}
		\end{center}
		
\begin{abstract}
	We construct an exact magnetized generalization of the Zipoy-Voorhees spacetime by applying the magnetic Harrison transformation to a static seed with quadrupolar deformation parameter $k$. The resulting Melvin-Zipoy-Voorhees metric is a solution to the Einstein-Maxwell equations that interpolates between the unmagnetized Zipoy-Voorhees geometry and the Melvin magnetic universe. We analyze the algebraic structure, finding the spacetime to be generically of Petrov type I, and investigate the equatorial dynamics of charged test particles and photons. Our analysis reveals that the external magnetic field $b$ induces a ``Lorentz shift'' in the effective angular momentum, suppressing the potential barrier and causing the Innermost Stable Circular Orbit (ISCO) to migrate inward. In contrast, the radius of the photon ring shifts slightly outward with increasing magnetization. 
\end{abstract}

	\end{titlepage}\onecolumn
	\bigskip
	
\section{Introduction}
\label{sec:intro}

The interplay between gravity and electromagnetic fields provides a fundamental framework for studying compact objects and their astrophysical environments \cite{Ernst1968,Newman1965,ErnstWild1976,FrolovShoom2010}. Among the most valuable theoretical settings are exact Einstein-Maxwell solutions describing static or stationary sources immersed in external magnetic fields. Such magnetized geometries, often generated through the Harrison transformation \cite{Harrison1968} within the Ernst formalism \cite{Ernst1968}, offer powerful laboratories for exploring curvature-electromagnetic coupling beyond spherical symmetry. Classic examples include the Schwarzschild-Melvin \cite{Melvin1964} and Kerr-Melvin \cite{ErnstWild1976} solutions, which model black holes embedded in a Melvin magnetic universe. These configurations have been extensively used to analyze energy extraction mechanisms, field line distortions, and magnetohydrodynamic environments around compact objects \cite{BlandfordZnajek1977, Nishikawa:2004wp, Komissarov:2006nz, Aliev:1989wz, Aliev:1989wx, Esteban:1988zp, Jumaniyozov:2025uwo}.

A natural extension of these studies involves considering deformed, non-spherical sources. The Zipoy-Voorhees spacetime \cite{Zipoy1966, Voorhees1970}---also known as the $\gamma$-metric---constitutes one of the simplest exact static, axisymmetric solutions of the Einstein equations that generalizes the Schwarzschild geometry through a real deformation parameter $k$. For $k=1$, spherical symmetry is recovered, while deviations ($k\neq1$) encode prolate or oblate quadrupolar distortions of the gravitational potential. Owing to its analytical simplicity and tunable deviation from Schwarzschild, the Zipoy-Voorhees metric has been employed in numerous contexts, ranging from fundamental geodesic studies \cite{Herrera:1998rj, Destounis:2023gpw} and accretion disk luminosity around compact objects with quadrupole moments \cite{Boshkayev:2021chc}, to specialized applications in string theory \cite{Yunusov:2025chw}. The framework has been further extended to include electromagnetic sources, with recent work on charged Zipoy-Voorhees spacetimes \cite{Toshmatov:2019bda,Toshmatov:2019qih, Benavides-Gallego:2018htf, Gurtug:2021noy} highlighting its role as a baseline for studying self-consistent electromagnetic fields and the influence of quadrupole deformations on particle dynamics and observational signatures.

Despite this progress, the magnetic counterpart of the Zipoy-Voorhees family has not been explicitly constructed in closed analytic form. Magnetizing non-spherical spacetimes presents both conceptual and technical challenges: the external field must be incorporated consistently within the Einstein-Maxwell system, and the resulting metric should preserve regularity outside the source while reducing smoothly to both the unmagnetized Zipoy-Voorhees and Melvin limits. The Harrison transformation \cite{Harrison1968}, formulated within the Ernst potential framework \cite{Ernst1968}, provides an elegant route to achieving this by embedding an arbitrary static seed solution into a uniform magnetic background.

In this work, we apply the magnetic Harrison transformation to the Zipoy-Voorhees seed and obtain an exact Melvin-Zipoy-Voorhees solution of the Einstein-Maxwell equations. The resulting configuration retains the quadrupolar deformation parameter $k$ while supporting an external magnetic field of strength $b$. The metric remains static and axisymmetric, reducing to the unmagnetized Zipoy-Voorhees geometry as $b \to 0$ and to the Melvin magnetic universe as the mass $M \to 0$ with $k \to 1$. We compute the associated electromagnetic field tensor and demonstrate that the field measured by static observers outside the compact object ($r > 2M$) is purely magnetic. Closed expressions are derived for the local components $B^{\hat{r}}$ and $B^{\hat{\theta}}$, illustrating how the deformation parameter $k$ modulates the angular distribution of magnetic flux.

We also analyze the curvature invariants and algebraic properties of the Melvin-Zipoy-Voorhees spacetime. The geometry is shown to be of Petrov type~I for generic parameters $(k,b)$, reducing to type~D only in the Schwarzschild limit $(k=1,b=0)$ \cite{Stephani2003}. The Kretschmann scalar exhibits smooth suppression under magnetization, with curvature profiles reflecting the coherent interplay between intrinsic quadrupolar deformation and the external magnetic energy density. In the asymptotic region, the metric approaches a Melvin-like structure, while along the symmetry axis it may display conical behavior analogous to other magnetized Weyl solutions \cite{Hiscock1985}, which can be regularized by appropriate azimuthal rescaling.

Furthermore, we perform a comprehensive analysis of equatorial dynamics for both massive charged particles and photons. By utilizing the Hamilton-Jacobi formalism, we demonstrate that the magnetic field induces a ``Lorentz shift'' in the effective angular momentum of charged test particles. This interaction modifies the centrifugal barrier, breaking the degeneracy between prograde and retrograde orbits and forcing the Innermost Stable Circular Orbit (ISCO) to migrate inward as magnetization increases. Crucially, we distinguish between the existence of equilibrium orbits and their physical viability by deriving the vertical stability criterion ($K_\theta > 0$), identifying regions where equatorial motion is unstable to perturbations orthogonal to the symmetry plane. In contrast to the inward shift of the ISCO, we find that the photon ring radius shifts outwardly with increasing magnetic field strength, offering distinct observational signatures for future shadow morphology studies \cite{Wang:2021ara}.

The organization of this paper is as follows. Section~\ref{sec:MZV} presents the derivation of the Melvin-Zipoy-Voorhees solution via the Ernst magnetization procedure. Section~\ref{sec:EM} analyzes the electromagnetic field structure and its fundamental properties, including curvature invariants and Petrov classification. The dynamics of massive charged test particles, with particular focus on circular orbital motion, stability, and the Lorentz shift, are examined in Section~\ref{sec:charged-test}. Section~\ref{sec:null-effective} investigates null geodesics and the behavior of the photon ring. We conclude in the last section. Throughout this work, we employ natural units where \( c = G_N = 1 \).
	
\section{Melvin-Zipoy-Voorhees Solution}\label{sec:MZV}

We take as our seed solution a static, axisymmetric spacetime that generalizes the Schwarzschild geometry through a deformation parameter $k$. The metric is constructed from the functions
\begin{equation}\label{eq.metricFunctions}
f(r) = 1 - \frac{2M}{r}, \qquad \Delta_r = r^2 - 2Mr, \qquad g(r,\theta) = 1 - \frac{2M}{r} + \frac{M^2 \sin^2\theta}{r^2},
\end{equation}
where $f(r)$ is the Schwarzschild lapse function and $g(r,\theta)$ encapsulates the angular deformation that generates a quadrupolar moment. The corresponding line element reads
\begin{equation}
\mathrm{d}s^2 = - f^{k}\, \mathrm{d}t^2 + f^{k^2 - k}\, g^{1 - k^2} \left( \frac{\mathrm{d}r^2}{f} + r^2\,\mathrm{d}\theta^2 \right) + f^{1 - k} r^2 \sin^2\theta\, \mathrm{d}\phi^2.
\end{equation}
This metric belongs to the Zipoy-Voorhees family and satisfies the vacuum Einstein equations. The case $k = 1$ recovers the spherical Schwarzschild solution, while values $k \neq 1$ yield a static, axially symmetric geometry with a nontrivial quadrupolar deformation.

To facilitate the magnetization procedure, we express the Zipoy-Voorhees metric in Lewis-Papapetrou-Weyl form:
\begin{equation}
\mathrm{d}s^2  = {\mathcal F}_0\big( \mathrm{d}\phi - \omega_0 \, \mathrm{d}t \big)^2  - \frac{\rho ^2 }{{\mathcal F}_0}\,\mathrm{d}t^2  + \frac{\mathrm{e}^{2\gamma } }{{\mathcal F}_0}\left( \frac{\mathrm{d}r^2 }{\Delta _r } + \mathrm{d}\theta^2 \right),
\end{equation}
where $\omega_0 = 0$, $\rho^2 = \Delta_r \sin^2\theta$, and
\begin{equation}
{\mathcal F}_0 = f^{1 - k}\, r^2 \sin^2\theta,
\end{equation}
\begin{equation}
\mathrm{e}^{2\gamma } = r^2 \left( \frac{r - 2M}{r} \right)^{(1 - k)^2}
\left( (r - M)^2 - M^2 \cos^2\theta \right)
\left( \frac{(r - M)^2 - M^2 \cos^2\theta}{r^2} \right)^{- k^2} \sin^2\theta.
\end{equation}

To implement the magnetic Harrison transformation with the standard Melvin normalization, we require the Ernst gravitational potential for the Zipoy-Voorhees spacetime:
\begin{equation}\label{eq.E0}
\mathcal{E}_0 = r^{2} \sin^{2}\theta\, f^{1-k},
\end{equation}
with vanishing electromagnetic Ernst potential, $\Phi_0 = 0$. Defining
\begin{equation}\label{eq.Lambda}
\Lambda \equiv 1 + \frac{b^{2}}{4}\, \mathcal{E}_0,
\end{equation}
we obtain the magnetized Ernst potentials
\begin{equation}
\mathcal{E} = \frac{\mathcal{E}_0}{\Lambda^{2}}, \qquad \Phi = \frac{b\, \mathcal{E}_0}{2\Lambda}.
\end{equation}

Using these potentials, the magnetized metric takes the form
\begin{equation}
\label{eq:magZV_metric}
\mathrm{d}s^{2} =
- \frac{\rho^{2}}{\mathcal{F}_b}\, \mathrm{d}t^{2}
+ \frac{\mathrm{e}^{2\gamma}}{\mathcal{F}_b} \left( \frac{\mathrm{d}r^{2}}{\Delta_r} + \mathrm{d}\theta^{2} \right) + \mathcal{F}_b \mathrm{d}\phi^{2},
\end{equation}
where 
\begin{equation}
\mathcal{F}_b = \frac{\mathcal{E}_0}{\Lambda^{2}}
= \frac{r^{2} f^{1-k} \sin^{2}\theta}{
	\big( 1 + \tfrac{b^{2}}{4} r^{2} f^{1-k} \sin^{2}\theta \big)^{2}}.
\end{equation}
The vector potential is purely azimuthal,
\begin{equation}
\label{eq:harrison_A}
A_\mu\, \mathrm{d}x^\mu = A_\phi\, \mathrm{d}\phi,
\qquad
A_\phi = \Phi = \frac{b\, r^{2} f^{1-k} \sin^{2}\theta}{2\Lambda}.
\end{equation}
These fields satisfy the Einstein-Maxwell equations
\begin{equation}
R_{\mu\nu} = 2 F_{\mu\alpha} F_{\nu}{}^{\alpha} - \frac{1}{2} g_{\mu\nu} F_{\alpha\beta} F^{\alpha\beta},
\end{equation}
and the source-free Maxwell equations
\begin{equation}
\nabla_\mu F^{\mu\nu} = 0, \qquad F_{\mu\nu} = \partial_\mu A_\nu - \partial_\nu A_\mu.
\end{equation}

\begin{figure}[bhtp]
	\centering
	\includegraphics[width=0.6\linewidth]{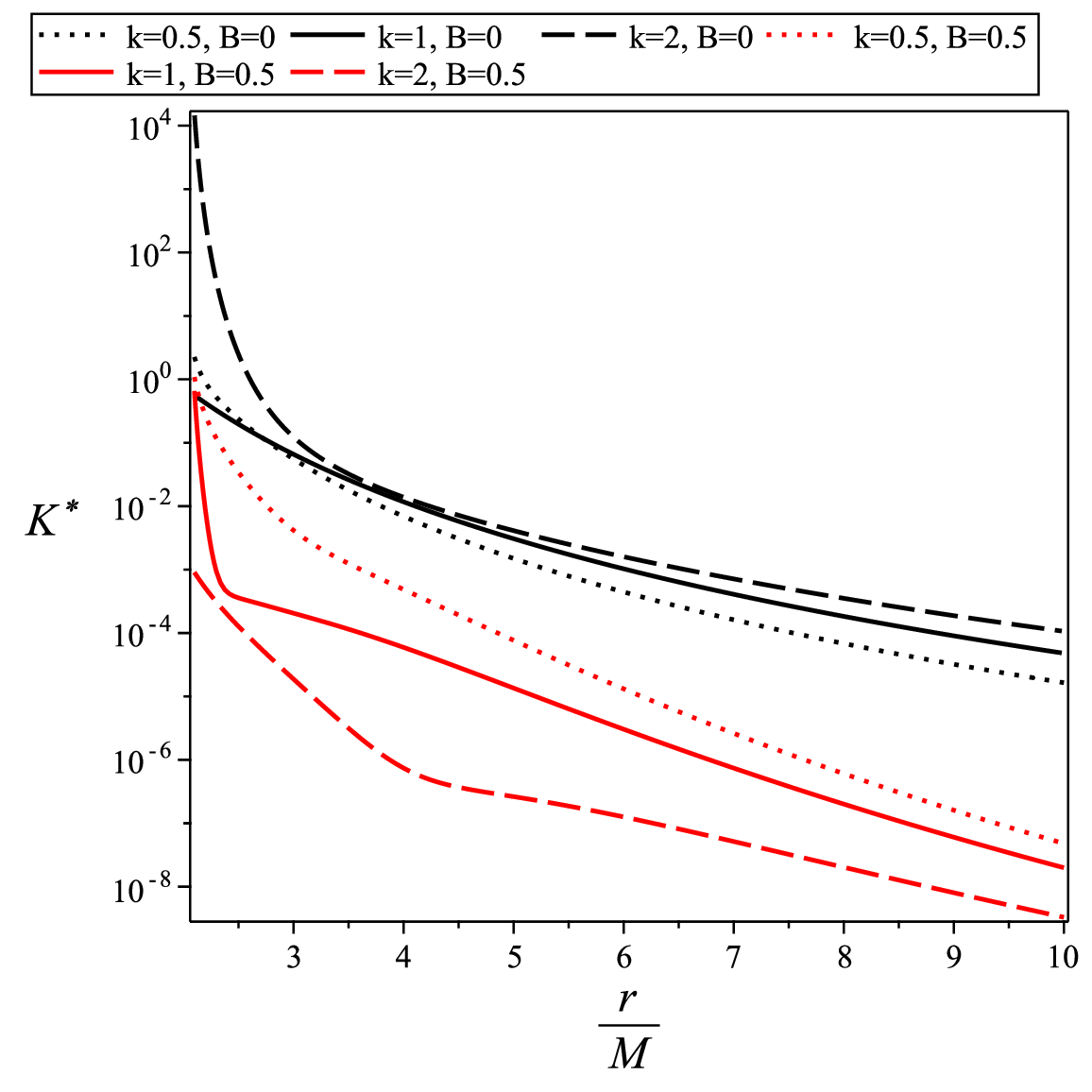}
	\caption{Logarithmic-linear plot of the dimensionless squared Riemann tensor $K^* = K M^4$
		on the equatorial plane of the Melvin-Zipoy-Voorhees spacetime for deformation parameters $k \in \{0.5, 1, 2\}$ and magnetic field strengths $b M \in \{0, 1\}$.
		Black curves correspond to $b = 0$; red curves to $b M = 1$.
		Line styles denote different deformation parameters: $k = 0.5$ (dotted), $k = 1$ (solid), $k = 2$ (dash-dotted).
		For fixed radial coordinate $r$, the curvature $K$ increases with $k$, while magnetization ($b M = 1$) suppresses the equatorial curvature amplitude throughout this radial range.
		All curves exhibit monotonic decay with increasing $r$.}
	\label{fig:K_equator}
\end{figure}

The squared Riemann tensor, or Kretschmann scalar $K \equiv R_{\mu\nu\rho\sigma} R^{\mu\nu\rho\sigma}$, quantifies the local energy density of the gravitational field. Figure~\ref{fig:K_equator} illustrates its radial dependence on the equatorial plane, demonstrating the combined influence of the magnetic field parameter $b$ and deformation parameter $k$. For the unmagnetized cases (black curves, $b = 0$), the curvature increases monotonically with $k$, confirming that stronger quadrupolar deformation enhances the curvature near the source. Crucially, the introduction of a magnetic field ($b M = 1$, red curves) suppresses the equatorial curvature amplitude across all deformation parameters in the depicted radial range, indicating that the electromagnetic field energy density partially counteracts the geometric curvature near the source. All profiles decay monotonically as $r \to \infty$, approaching the asymptotic behavior of the background spacetime.

For completeness, we summarize the algebraic classification procedure used to determine the Petrov type of the Melvin-Zipoy-Voorhees spacetime. Let $C_{\alpha\beta\gamma\delta}$ denote the Weyl tensor, and
\begin{equation}
{}^\star C_{\alpha\beta\gamma\delta}
= \tfrac{1}{2}\,\varepsilon_{\alpha\beta}{}^{\mu\nu}C_{\mu\nu\gamma\delta}
\end{equation}
its Hodge dual, where $\varepsilon_{\alpha\beta\gamma\delta}$ is the totally antisymmetric Levi-Civita tensor in the metric signature $(-+++)$. The complex, self-dual Weyl tensor is then defined as \cite{Stephani2003,Penrose1984}
\begin{equation}
\tilde C_{\alpha\beta\gamma\delta}
= C_{\alpha\beta\gamma\delta}
+ i\,{}^\star C_{\alpha\beta\gamma\delta}.
\label{eq:Csd}
\end{equation}
This tensor acts naturally on the three-dimensional complex vector space of self-dual bivectors, and its algebraic properties determine the Petrov classification \cite{Stephani2003,Chandrasekhar1983}.

To classify the spacetime, we compute the complex scalar invariants \cite{Stephani2003,MacCallum2006,Podolsky2009}
\begin{equation}
I = \tfrac{1}{8}\,\tilde C_{\alpha\beta\gamma\delta}
\tilde C^{\alpha\beta\gamma\delta},\qquad
J = \tfrac{1}{48}\,\tilde C_{\alpha\beta}{}^{\gamma\delta}
\tilde C_{\gamma\delta}{}^{\mu\nu}
\tilde C_{\mu\nu}{}^{\alpha\beta}.
\end{equation}
These invariants were calculated from the Weyl scalars in a null tetrad to ensure standard Petrov normalization. In the Schwarzschild limit $(k=1, b=0)$, we find $27J^2 = I^3$ exactly, as required for a Petrov type~D spacetime \cite{Chandrasekhar1983}. For generic parameter values $(k, b)$, however, the discriminant $T_p \equiv 27J^2 - I^3$ is non-vanishing, confirming that the general Melvin-Zipoy-Voorhees spacetime is of Petrov type~I \cite{Stephani2003}.

\section{Electromagnetic properties in the spacetime}\label{sec:EM}

In curved spacetime, the electromagnetic field is described by the antisymmetric field strength tensor
\begin{equation}
F_{\mu\nu}=\partial_\mu A_\nu-\partial_\nu A_\mu .
\end{equation}
The magnetic field measured by a local observer with 4-velocity \(u^\mu\) is obtained by projecting \(F_{\mu\nu}\) onto the observer's orthonormal tetrad \(e^{\hat\mu}{}_\nu\). The components of the magnetic field 3-vector in this local frame are
\begin{equation}
B^{\hat\mu}=\frac{1}{2}\,\epsilon^{\hat\mu\hat\nu\hat\alpha\hat\beta}\,u_{\hat\nu}\,F_{\hat\alpha\hat\beta},
\end{equation}
where \(\epsilon^{\hat\mu\hat\nu\hat\alpha\hat\beta}\) is the totally antisymmetric Levi-Civita symbol in the orthonormal frame, and \(F_{\hat\mu\hat\nu}=e_{\hat\mu}{}^{\alpha}\,e_{\hat\nu}{}^{\beta}\,F_{\alpha\beta}\) are the frame components.

For a static observer, \(u^\mu=\delta^\mu_t/\sqrt{-g_{tt}}\), in a static axisymmetric spacetime with purely magnetic potential \(A_\mu=A_\phi(r,\theta)\,\delta^\phi_\mu\), the non-zero coordinate components are
\begin{equation}
F_{r\phi}=\partial_r A_\phi,\qquad F_{\theta\phi}=\partial_\theta A_\phi .
\end{equation}
Consistently with the metric \eqref{eq:magZV_metric} and potential \eqref{eq:harrison_A}, an orthonormal basis of one-forms for static observers is
\begin{equation}\label{eq:OrthoStatic}
\begin{aligned}
e^{\hat t}&=\sqrt{-g_{tt}}\,\mathrm{d}t
= f^{k/2}\Lambda\,\mathrm{d}t,\\[2pt]
e^{\hat r}&=\sqrt{g_{rr}}\,\mathrm{d}r
=\frac{\mathrm{e}^{\gamma}\Lambda}{\sqrt{\mathcal{E}_0\,\Delta_r}}\,\mathrm{d}r,\\[2pt]
e^{\hat\theta}&=\sqrt{g_{\theta\theta}}\,\mathrm{d}\theta
=\frac{\mathrm{e}^{\gamma}\Lambda}{\sqrt{\mathcal{E}_0}}\,\mathrm{d}\theta,\\[2pt]
e^{\hat\phi}&=\sqrt{g_{\phi\phi}}\,\mathrm{d}\phi
=\frac{\sqrt{\mathcal{E}_0}}{\Lambda}\,\mathrm{d}\phi,
\end{aligned}
\end{equation}
where the functions are given in Eqs. (\ref{eq.metricFunctions}), (\ref{eq.Lambda}), and (\ref{eq.E0}).

Using \(A_\phi=(b/2)\,\mathcal{E}_0/\Lambda\) and the identity
\begin{equation}
\partial_j\!\left(\frac{\mathcal{E}_0}{\Lambda}\right)
=\frac{\partial_j \mathcal{E}_0}{\Lambda^2},\qquad j\in\{r,\theta\},
\end{equation}
the non-zero components of \(F_{\mu\nu}\) are
\begin{equation}
F_{r\phi}=\frac{b}{2\Lambda^2}\,\partial_r\mathcal{E}_0,\qquad
F_{\theta\phi}=\frac{b}{2\Lambda^2}\,\partial_\theta\mathcal{E}_0 .
\end{equation}
Projecting to the static orthonormal frame yields the measured magnetic field
\begin{equation}\label{eq:BhatCOMP}
\begin{aligned}
B^{\hat r}
&=\frac{F_{\theta\phi}}{\sqrt{g_{\theta\theta}g_{\phi\phi}}}
=\frac{F_{\theta\phi}}{\mathrm{e}^{\gamma}}
=\frac{b\,r^{2} f^{\,1-k}\,\sin\theta\cos\theta}{\Lambda^{2}\,\mathrm{e}^{\gamma}},\\[4pt]
B^{\hat\theta}
&=\frac{F_{\phi r}}{\sqrt{g_{rr}g_{\phi\phi}}}
=-\,\frac{F_{r\phi}}{\mathrm{e}^{\gamma}\sqrt{\Delta_r}}
=-\,\frac{b\,\sqrt{\Delta_r}}{\Lambda^{2}\,\mathrm{e}^{\gamma}}\,
\sin^{2}\theta\,
\Big(r f^{\,1-k}+M(1-k)f^{-k}\Big),\\[4pt]
B^{\hat\phi}&=0,
\end{aligned}
\end{equation}
which are manifestly real for \(r>2M\) and \(0<\theta<\pi\). On the axes \((\theta=0,\pi)\) one has \(B^{\hat\theta}=0\) and \(B^{\hat r}\to 0\) linearly with \(\sin\theta\); on the equator \((\theta=\pi/2)\), \(B^{\hat r}=0\) and \(B^{\hat\theta}\) carries the magnetic energy density.

The Maxwell field is characterized by two invariants,
\begin{equation}
I_1\equiv F_{\mu\nu}F^{\mu\nu},\qquad
I_2\equiv F_{\mu\nu}\,{}^{\star}\!F^{\mu\nu},
\end{equation}
with \({}^{\star}F_{\mu\nu}=\tfrac12\varepsilon_{\mu\nu}{}^{\rho\sigma}F_{\rho\sigma}\). In the static frame \(\mathbf{E}=0\), so $I_2=0$ and
\begin{equation}\label{eq:invariantsPos}
I_1=2\mathbf{B}^{2}
=2\left[(B^{\hat r})^{2}+(B^{\hat\theta})^{2}\right]\ge 0,
\end{equation}
for \(r>2M\). In the limit \(M\to0\) and \(k\to1\) one has \(f\to1\), \(\mathrm{e}^{\gamma}\to r^{2}\sin\theta\), \(\Delta_r\to r^{2}\), and \(\Lambda\to 1+\tfrac{b^{2}}{4}r^{2}\sin^{2}\theta\), so
\begin{equation}
B^{\hat r}\to \frac{b\,\cos\theta}{\Lambda^{2}},\qquad
B^{\hat\theta}\to -\,\frac{b\,\sin\theta}{\Lambda^{2}},\qquad
B^{\hat\phi}=0,
\end{equation}
and therefore
\begin{equation}
I_1\to 2\left[(B^{\hat r})^{2}+(B^{\hat\theta})^{2}\right]
=\frac{2b^{2}}{\Lambda^{4}},\qquad I_2\to0,
\end{equation}
which is the Melvin behavior in this gauge.

The magnetic field exhibits a characteristic angular pattern inherited from the magnetic Harrison transform acting on a static Zipoy-Voorhees seed: \(A_\phi=b\,\mathcal{E}_0/\Lambda=b\,r^{2}f^{\,1-k}\sin^{2}\theta/\Lambda\) produces non-vanishing \(F_{r\phi}=\partial_r A_\phi\) and \(F_{\theta\phi}=\partial_\theta A_\phi\) but no electric component in the static frame. Consequently \(E_{\hat i}=0\), \(I_2=0\), and \(I_1=2\mathbf{B}^{2}\ge0\) as in \eqref{eq:invariantsPos}. From \eqref{eq:BhatCOMP} one sees that \(B^{\hat r}\propto \sin\theta\cos\theta\) and \(B^{\hat\theta}\propto \sin^{2}\theta\big(r f^{\,1-k}+M(1-k)f^{-k}\big)\), reproducing the Melvin-like pattern in the limit \(k\to1\), \(M\to0\), while making explicit how the Zipoy-Voorhees deformation \(k\) reweights the near-source field through powers of \(f(r)\). For fixed \(r/M\) and \(b\), increasing \(k\) redistributes magnetic flux between polar and equatorial regions via the \(k\)-dependent factors in \eqref{eq:BhatCOMP}, without introducing any electric component. All statements hold for the exterior region \(r>2M\).

\begin{figure}[bhtp]
	\centering
	\includegraphics[width=0.6\linewidth]{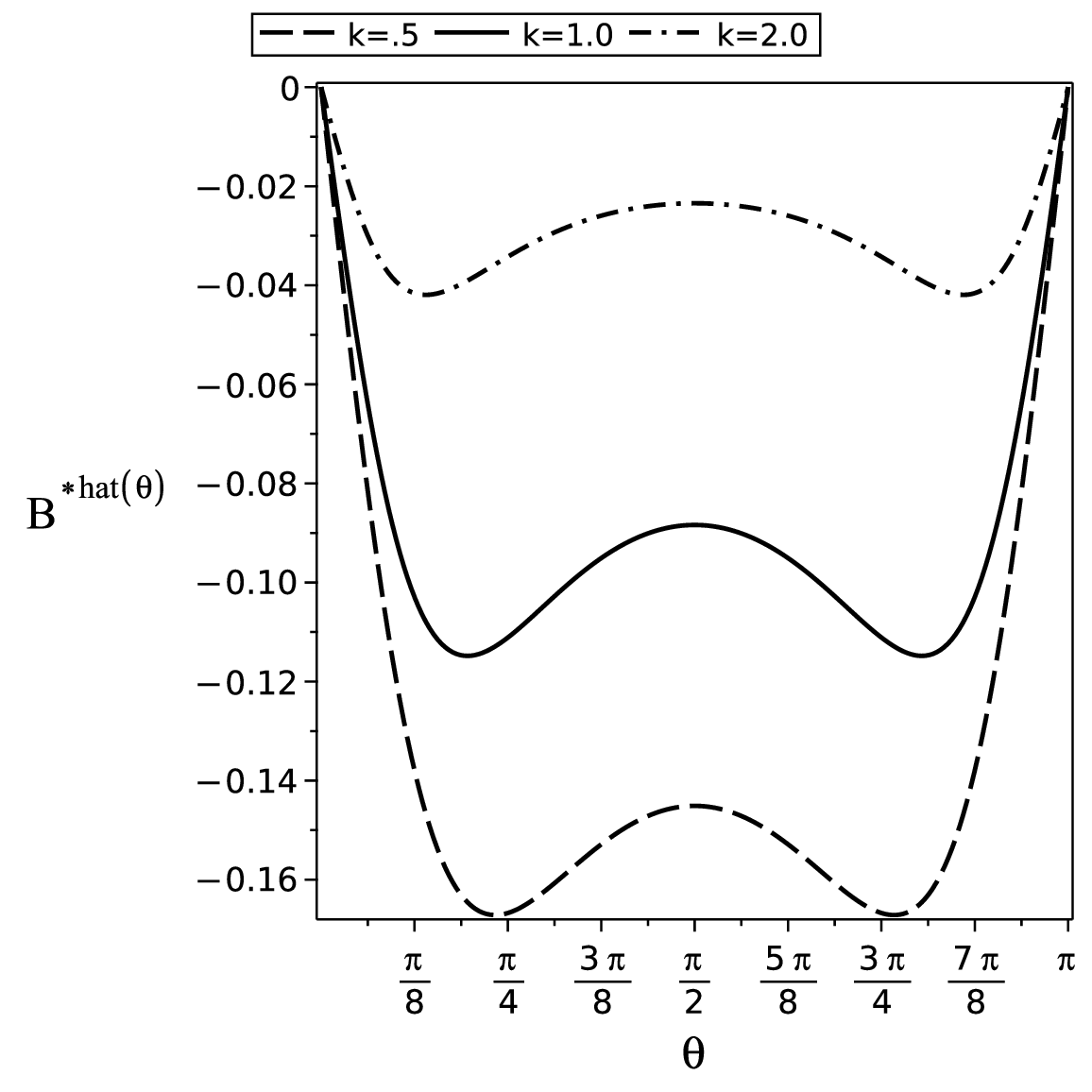}
	\caption{Polar magnetic field measured by a static observer.
		Angular profiles of dimensionless $B^{*\hat\theta}(\theta)=M B^{\hat\theta}(\theta)$ at fixed radius $r=4M$ for
		$k\in\{0.5,1,2\}$ with $bM=0.5$. The field is evaluated in the static orthonormal frame
		[Eq.~\eqref{eq:OrthoStatic}] using the Harrison-magnetized potential
		[Eq.~\eqref{eq:harrison_A}]. From Eq.~\eqref{eq:BhatCOMP} the profile is symmetric about the equator and vanishes on the axis.}
	\label{fig:Bth_slice}
\end{figure}

\begin{figure}[bhtp]
	\centering
	\includegraphics[width=0.6\linewidth]{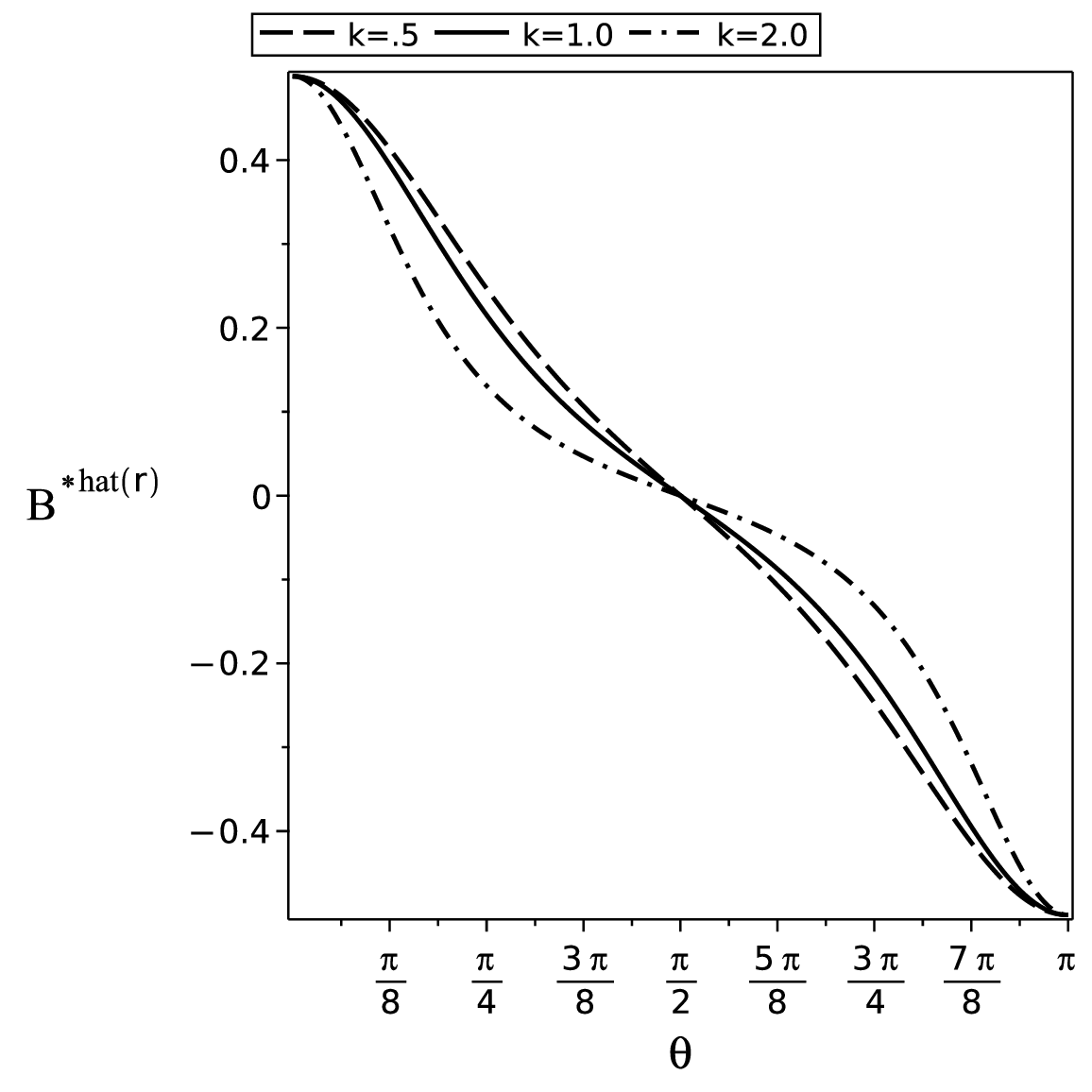}
\caption{Radial magnetic field measured by a static observer.
	Angular profiles of dimensionless $B^{*\hat r}(\theta)=M B^{\hat r}(\theta)$ at fixed radius $r=3M$ for
	$k\in\{0.5,1,2\}$ with $bM=0.5$. The field is evaluated in the static orthonormal frame
	[Eq.~\eqref{eq:OrthoStatic}] using the Harrison-magnetized potential
	[Eq.~\eqref{eq:harrison_A}]. From Eq.~\eqref{eq:BhatCOMP}, $B^{\hat r}$ is antisymmetric about the equator and vanishes at $\theta=\pi/2$, while remaining finite and of opposite sign on the two axes. Its amplitude is modulated by $k$ through the factors $f^{1-k}$, $\Lambda^{-2}$, and $\mathrm{e}^{-\gamma}$.}
	\label{fig:Br_slice}
\end{figure}

\begin{figure*}[bhtp]
	\centering
	\begin{subfigure}[t]{0.47\textwidth}
		\centering
		\includegraphics[width=\linewidth]{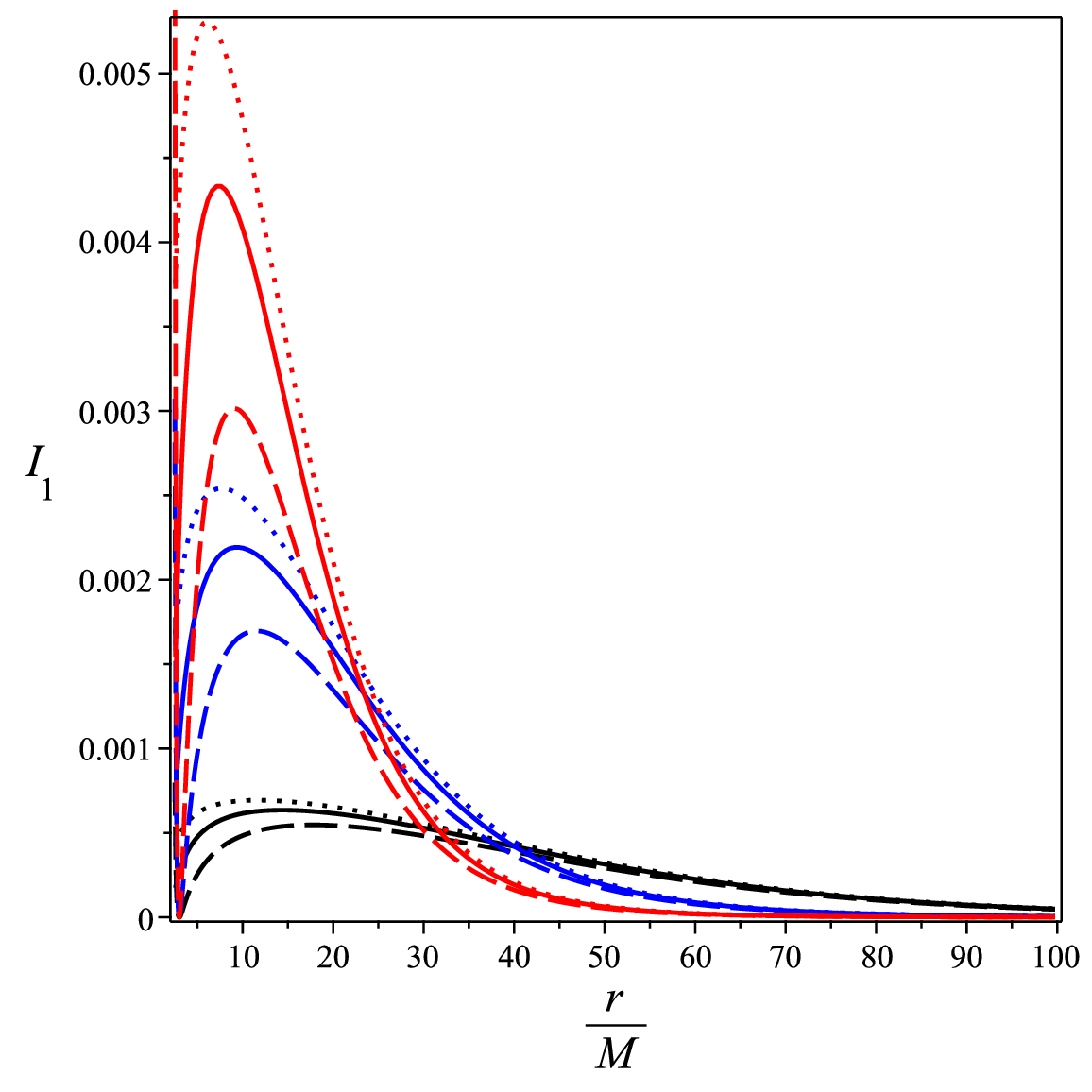}
		\caption{Near-source region ($r/M\in[2,2.5]$), logarithmic vertical axis.
			Equatorial profiles of the electromagnetic invariant
			$I_1=F_{\mu\nu}F^{\mu\nu}$ for $k\in\{0.5,1,2\}$ and
			$bM\in\{0.02,0.04,0.06\}$.
			Colors indicate $bM$ (black: 0.02, blue: 0.04, red: 0.06);
			line styles represent $k$ (dotted: 0.5, solid: 1, dash-dot: 2).
			The logarithmic scale resolves the steep growth just outside $r=2M$ while keeping all curves visible.}
		\label{fig:I1_near}
	\end{subfigure}\hfill
	\begin{subfigure}[t]{0.47\textwidth}
		\centering
		\includegraphics[width=\linewidth]{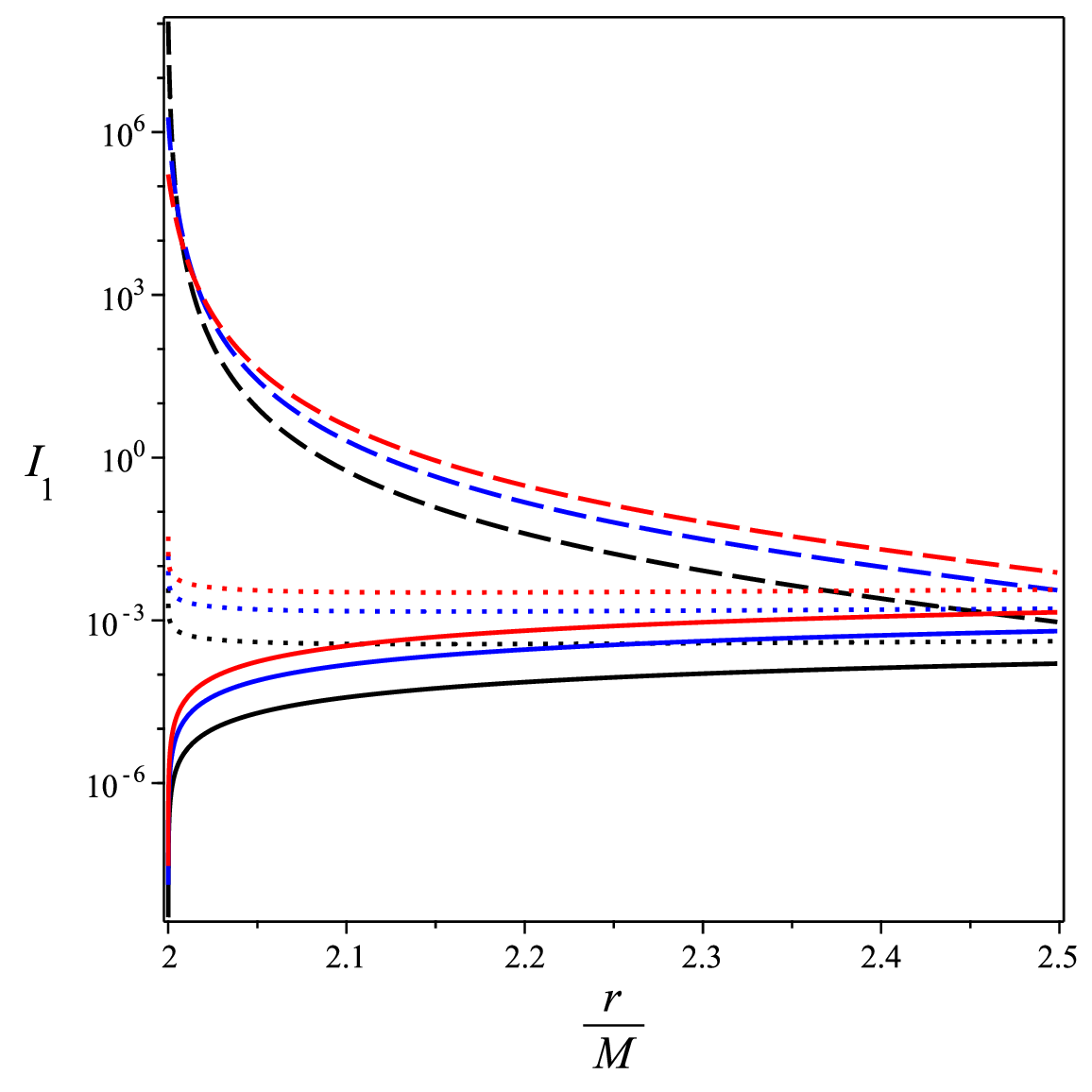}
		\caption{Exterior decay ($r/M\in[2.5,100]$), linear vertical axis.
			The same configurations as in panel (a), showing a single exterior peak (typically at $r/M\sim\mathcal{O}(10)$) followed by monotonic decay.
			Larger $bM$ increases the overall amplitude, consistent with the small-$b$ scaling $I_1\propto b^{2}$, while larger $k$ enhances the near-source field and shifts the peak position.}
		\label{fig:I1_far}
	\end{subfigure}
	\caption{Electromagnetic invariant on the equatorial plane.
		We plot $I_1(r,\theta=\pi/2)$ for the Melvin-Zipoy-Voorhees spacetime, where $I_1=2\mathbf{B}^2\ge0$ for static observers [cf. Eq.~\eqref{eq:invariantsPos}].
		Panel (a) uses a logarithmic axis to resolve the sharp growth just outside $r=2M$; panel (b) displays the full exterior decay on a linear scale.}
	\label{fig:I1_equator_panels}
\end{figure*}

\section{Circular motions of massive charged test particles}
\label{sec:charged-test}

In this section, we study the equatorial circular motion of massive charged test particles in the Melvin--Zipoy--Voorhees spacetime using the Hamilton--Jacobi formalism. The principal function $S$ satisfies the relativistic Hamilton--Jacobi equation coupled to the electromagnetic field,
\begin{equation}
g^{\mu\nu}\bigl(\partial_\mu S-qA_\mu\bigr)\bigl(\partial_\nu S-qA_\nu\bigr)=-\mu^2,
\label{eq:HJ}
\end{equation}
where $\mu$ and $q$ are the particle's rest mass and electric charge, respectively. The stationarity and axisymmetry of the spacetime imply the existence of conserved quantities associated with the Killing vectors $\xi^\alpha=(\partial_t)^\alpha$ and $\psi^\alpha=(\partial_\phi)^\alpha$. Restricting our analysis to the equatorial plane $\theta=\pi/2$ (where $\dot\theta=0$ is consistent with reflection symmetry), we adopt the standard ansatz for the action,
\begin{equation}
S=-Et+L_z\phi+S_r(r),
\label{eq:HJ-ansatz}
\end{equation}
where the constants of motion are the energy $E=-p_t$ and the axial angular momentum $L_z=p_\phi$. On the equator, the relevant non-zero metric components and the vector potential are given by
\begin{equation}
g_{tt}=-f^{k}\Lambda^{2},\qquad
g_{\phi\phi}=\frac{r^{2}f^{1-k}}{\Lambda^{2}},\qquad
A_\phi=\frac{b\,r^{2}f^{1-k}}{2\Lambda},
\label{eq:eq-metric-Aphi}
\end{equation}
where $b$ is the external magnetic field parameter.

It is convenient to work with quantities normalized per unit rest mass. We define the specific energy $\mathcal E\equiv E/\mu$, the specific angular momentum $\mathcal L\equiv L_z/\mu$, and the charge-to-mass ratio $\eta\equiv q/\mu$. The dynamics are governed by the gauge-invariant (mechanical) azimuthal angular momentum per unit mass,
\begin{equation}
\tilde{\mathcal L}\equiv \mathcal L-\eta A_\phi
=\mathcal L-\eta\,\frac{b\,r^{2}f^{1-k}}{2\Lambda}.
\label{eq:Ltilde}
\end{equation}
The azimuthal equation of motion can then be expressed as
\begin{equation}
\dot\phi=\frac{L_z-qA_\phi}{\mu\,g_{\phi\phi}}
=\frac{\tilde{\mathcal L}}{g_{\phi\phi}}
=\frac{\Lambda^{2}}{r^{2}f^{1-k}}\tilde{\mathcal L}.
\end{equation}

Substituting the ansatz \eqref{eq:HJ-ansatz} into \eqref{eq:HJ} and specializing to equatorial motion yields the radial equation
\begin{equation}
g^{rr}\Bigl(\frac{dS_r}{dr}\Bigr)^2+g^{tt}E^2
+g^{\phi\phi}\bigl(L_z-qA_\phi\bigr)^2+\mu^2=0.
\end{equation}
Dividing by $\mu^2$ and defining the normalized radial action $\hat S_r\equiv S_r/\mu$, we obtain the compact form
\begin{equation}
g^{rr}\Bigl(\frac{d\hat S_r}{dr}\Bigr)^2
+g^{tt}\,\mathcal E^2
+g^{\phi\phi}\,\tilde{\mathcal L}^{\,2}
+1=0.
\label{eq:HJ-radial-compact}
\end{equation}
For timelike trajectories, this equation is equivalent to a one-dimensional radial problem,
\begin{equation}
\mathcal R(r)\equiv \mathcal E^2-U_{\rm eff}(r;\mathcal L, \eta),
\end{equation}
where the radial function $\mathcal R(r)$ is related to the radial momentum by
\begin{equation}
\left( \frac{d\hat{S}_r}{dr} \right)^2 = \frac{g_{rr}}{(-g_{tt})} \mathcal{R}(r).
\end{equation}
From this relation, it is clear that physically allowed regions require $\mathcal R(r) \ge 0$, and turning points occur where $\mathcal R(r)=0$. The effective potential $U_{\rm eff}$ for charged particle motion is given by
\begin{equation}
U_{\rm eff}(r;\mathcal L, \eta)
= (-g_{tt})\!\left(1+\frac{\tilde{\mathcal L}^{\,2}}{g_{\phi\phi}}\right)
= f^{k}\Lambda^{2}\!\left[1+\frac{\Lambda^{2}}{r^{2}f^{1-k}}\left(\mathcal L - \eta A_\phi \right)^{\!2}\right].
\label{eq:Radial_Eq_Ueff}
\end{equation}
Substituting the explicit form of $A_\phi$ from Eq.~\eqref{eq:eq-metric-Aphi}, the potential becomes
\begin{equation}
U_{\rm eff}(r;\mathcal L, \eta) = f^{k}\Lambda^{2}\!\left[1+\frac{\Lambda^{2}}{r^{2}f^{1-k}}\left(\mathcal L - \eta\,\frac{b\, r^{2} f^{1-k}}{2\Lambda}\right)^{\!2}\right].
\end{equation}

Explicitly, the radial function $\mathcal R(r)$ takes the algebraic form
\begin{equation}
\mathcal R(r)=\mathcal E^{2}+\mathcal A(r)\left[1+\frac{\mathcal B(r)^{2}}{\mathcal C(r)}\right],
\label{eq:R-explicit}
\end{equation}
where the auxiliary functions are defined as
\begin{align}
\mathcal A(r)=\;&
4rb^{2}
+4\Bigl(\frac{r-2}{r}\Bigr)^{-k}r^{3}b^{4}
-4\Bigl(\frac{r-2}{r}\Bigr)^{-k}r^{2}b^{4}
-\Bigl(\frac{r-2}{r}\Bigr)^{k}
-2r^{2}b^{2}
-\Bigl(\frac{r-2}{r}\Bigr)^{-k}r^{4}b^{4},\\[2mm]
\mathcal B(r)=\;&
\eta\,\frac{\Bigl(\frac{r-2}{r}\Bigr)^{1-k}r^{2}b}{\Bigl(\frac{r-2}{r}\Bigr)^{1-k}r^{2}b^{2}+1}
+{\mathcal L},\\[2mm]
\mathcal C(r)=\;&
\frac{\Bigl(\frac{r-2}{r}\Bigr)^{k}}{r(r-2)}
\left(r^{2}b^{2}-2rb^{2}+\Bigl(\frac{r-2}{r}\Bigr)^{k}\right)^{2}.
\end{align}

\begin{figure}[hbtp]
	\centering
	\includegraphics[width=0.6\linewidth]{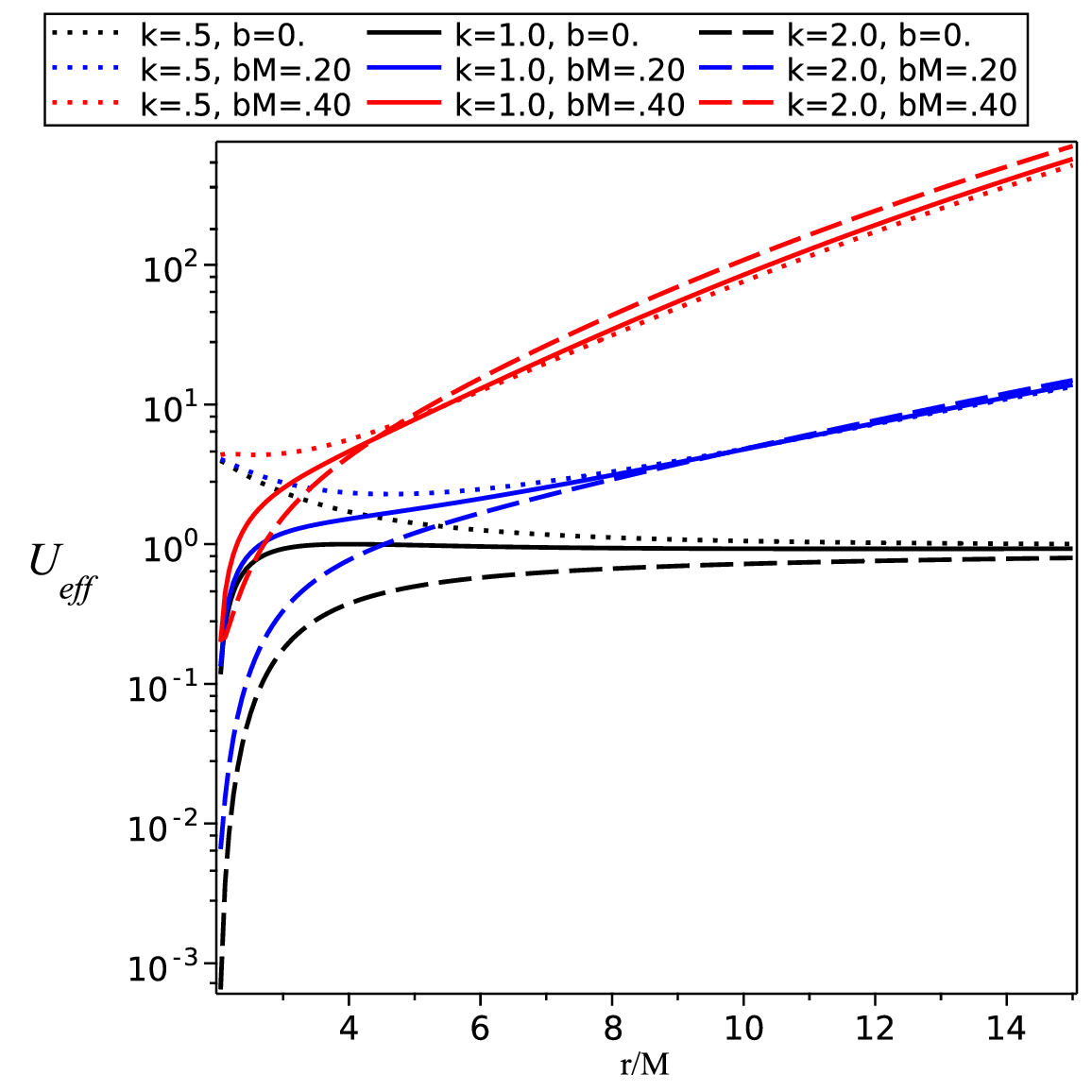} 
	\caption{Equatorial effective potential on a semi-log scale. 
		Shown is $U_{\rm eff}(r; L_z, \eta)$ for fixed angular momentum $L_z=4$ and charge parameter $\eta=0.10$.
		Curves vary the deformation parameter $k\in\{0.5,1,2\}$ (line style) and the magnetic parameter $bM\in\{0,0.20,0.40\}$ (color).
		The black curves ($b=0$) show the unmagnetized Zipoy--Voorhees reference. 
		Increasing $bM$ (blue and red curves) significantly lowers the potential barrier, a direct consequence of the electromagnetic interaction term $-\eta A_\phi$ shifting the effective angular momentum (the Lorentz shift discussed in Eq.~\eqref{eq:Radial_Eq_Ueff}).}
	\label{fig:Ueff_kB_fixedLz}
\end{figure}

The ISCO is defined by the standard marginal-stability conditions
\begin{equation}
\mathcal R(r_c)=0,\qquad \mathcal R'(r_c)=0,\qquad \mathcal R''(r_c)=0,
\label{eq:ISCOconds}
\end{equation}
where the prime denotes differentiation with respect to $r$. These three equations determine the triplet $(r_c,\mathcal E,\mathcal L)$ on each orbital branch. To investigate the ISCO behavior, we solve the system~\eqref{eq:ISCOconds} numerically for representative deformation parameters $k\in\{\tfrac12,1,2\}$, fixing the charge-to-mass ratio at $\eta=0.1$ and working in the weak-field regime by truncating the magnetization dependence at ${\cal O}(b^2)$. Solving~\eqref{eq:ISCOconds} with the full (all-orders) magnetized expressions becomes cumbersome, whereas the ${\cal O}(b^2)$ truncation is physically well-motivated since any astrophysical magnetic field threading the exterior region is expected to satisfy $bM\ll 1$.

The resulting ISCO radii are reported in Tables~\ref{tab:isco-k2}--\ref{tab:isco-khalf}, separated into prograde ($\mathcal L>0$) and retrograde ($\mathcal L<0$) branches. For each listed solution, we verified the stability assignment by checking the sign of $\mathcal R''$ in the vicinity of $r_c$: circular orbits just outside the ISCO ($r=r_c+\epsilon$) satisfy $\mathcal R''(r)<0$ (local minimum of $U_{\rm eff}$), while the corresponding inner branch ($r=r_c-\epsilon$) has $\mathcal R''(r)>0$ and is unstable, where $\epsilon>0$ is taken sufficiently small.

The influence of the magnetic field on the radial stability of charged particles is illustrated in Figure~\ref{fig:Ueff_kB_fixedLz}, which depicts the equatorial effective potential $U_{\rm eff}(r)$ for fixed angular momentum $L_z=4$ and charge parameter $\eta=0.10$. The curves reveal the combined impact of the quadrupolar deformation parameter $k$ and the magnetic field strength $b$. A salient feature is the significant suppression of the potential barrier as the magnetic field increases from the vacuum case ($b=0$, black curves) to magnetized states ($bM=0.20$ and $0.40$, blue and red curves). This reduction is driven dynamically by the electromagnetic interaction term $-\eta A_\phi$, which induces a ``Lorentz shift'' in the particle's effective angular momentum. 

This suppression of the potential barrier has a direct dynamical consequence on orbital stability: it forces the ISCO to migrate inward. This trend is quantitatively confirmed in Tables~\ref{tab:isco-k2}--\ref{tab:isco-khalf}. Across all deformation regimes, an increase in the magnetic parameter $bM$ consistently reduces the ISCO radius $r_{\rm ISCO}$. For instance, in the highly prolate regime ($k=2$, Table~\ref{tab:isco-k2}), the ISCO contracts from $11.36M$ in the vacuum case to approximately $10.77M$ at $bM=0.005$. This inward shift occurs because the electromagnetic interaction effective deepens the potential well, allowing stable circular motion to persist at smaller radii than would be possible in a purely vacuum geometry. Furthermore, the magnetic coupling breaks the degeneracy between prograde (${\cal L}>0$) and retrograde (${\cal L}<0$) orbits, inducing a fine splitting in their respective ISCO radii that grows with the magnetic field strength.

The existence of equatorial orbits is guaranteed by the reflection symmetry of the magnetized spacetime. Specifically, at the equatorial plane ($\theta = \pi/2$), the first derivatives of both the metric components and the vector potential with respect to the poloidal angle vanish ($\partial_\theta g_{\mu\nu} = 0$ and $\partial_\theta A_\phi = 0$). Consequently, the generalized forces orthogonal to the plane disappear, ensuring that the Euler-Lagrange equations for circular motion are strictly obeyed at the equator. However, this condition alone does not guarantee that such orbits are physically realizable. The physical viability of equatorial motion depends critically on the stability of this equilibrium against vertical perturbations, governed by the sign of the second derivative $K_\theta \equiv \partial^2_\theta U_{\rm eff}|_{\pi/2}$. In the Schwarzschild limit ($k=1, b=0$), $K_\theta$ is strictly positive outside the horizon, providing a restoring force that confines particles to the orbital plane. In contrast, for the general magnetized Zipoy-Voorhees spacetime, the interplay between the quadrupolar deformation $k$ and the magnetic field $b$ modifies the spacetime curvature---augmented by the direct electromagnetic coupling between the test object and the external magnetic field---such that $K_\theta$ is not universally positive. Regions where $K_\theta < 0$ represent a potential ``ridge'' rather than a valley; in these zones, the equatorial motion is unstable, and any infinitesimal vertical perturbation will cause the test particle to be ejected into a non-equatorial trajectory. Consequently, the ISCO configurations presented in Tables~\ref{tab:isco-k2}--\ref{tab:isco-khalf} are stable under vertical ($\theta$) perturbations, as we have verified $K_\theta > 0$ for all listed parameter sets ${r_{\rm ISCO},{\cal L},{\cal E},b, k}$. It is crucial to distinguish that while $K_\theta < 0$ signals instability to perturbations off the equatorial plane, it does not negate the existence of the ISCO itself, which is defined strictly by radial stability criteria.

\begin{table}[bhtp]
	\centering
	\begin{tabular}{c c c c}
		\hline
		$bM$ & $r_{\rm ISCO}/M$ & ${\cal L}$ & ${\cal E}$ \\
		\hline
		\multicolumn{4}{c}{\textbf{Prograde} (${\cal L}>0$)} \\
		\hline
		0.000 & 11.35890 & $+7.01804$ & 0.94465 \\
		0.001 & 11.32741 & $+7.02935$ & 0.94571 \\
		0.002 & 11.23882 & $+7.05305$ & 0.94753 \\
		0.003 & 11.10729 & $+7.08823$ & 0.95009 \\
		0.004 & 10.94920 & $+7.13366$ & 0.95337 \\
		0.005 & 10.77862 & $+7.18804$ & 0.95731 \\
		\hline
		\multicolumn{4}{c}{\textbf{Retrograde} (${\cal L}<0$)} \\
		\hline
		0.000 & 11.35890 & $-7.01804$ & 0.94465 \\
		0.001 & 11.32723 & $-7.01951$ & 0.94438 \\
		0.002 & 11.23753 & $-7.03345$ & 0.94488 \\
		0.003 & 11.10345 & $-7.05901$ & 0.94611 \\
		0.004 & 10.94135 & $-7.09500$ & 0.94806 \\
		0.005 & 10.76557 & $-7.14014$ & 0.95067 \\
		\hline
	\end{tabular}	\caption{ISCO data for $k=2$ separated into prograde (${\cal L}>0$) and retrograde (${\cal L}<0$) branches.}
	\label{tab:isco-k2}
\end{table}

\begin{table}[bhtp]
	\centering
	\begin{tabular}{c c c c}
		\hline
		$bM$ & $r_{\rm ISCO}/M$ & ${\cal L}$ & ${\cal E}$ \\
		\hline
		\multicolumn{4}{c}{\textbf{Prograde} (${\cal L}>0$)} \\
		\hline
		0.000 & 6.00000 & $+3.46410$ & 0.94281 \\
		0.001 & 5.99654 & $+3.46601$ & 0.94323 \\
		0.002 & 5.98634 & $+3.46933$ & 0.94383 \\
		0.003 & 5.96985 & $+3.47402$ & 0.94460 \\
		0.004 & 5.94772 & $+3.48006$ & 0.94556 \\
		0.005 & 5.92076 & $+3.48738$ & 0.94669 \\
		\hline
		\multicolumn{4}{c}{\textbf{Retrograde} (${\cal L}<0$)} \\
		\hline
		0.000 & 6.00000 & $-3.46410$ & 0.94281 \\
		0.001 & 5.99653 & $-3.46361$ & 0.94257 \\
		0.002 & 5.98627 & $-3.46453$ & 0.94252 \\
		0.003 & 5.96960 & $-3.46684$ & 0.94264 \\
		0.004 & 5.94716 & $-3.47051$ & 0.94295 \\
		0.005 & 5.91973 & $-3.47547$ & 0.94342 \\
		\hline
	\end{tabular}	\caption{ISCO data for $k=1$ separated into prograde (${\cal L}>0$) and retrograde (${\cal L}<0$) branches.}
	\label{tab:isco-k1}
\end{table}

\begin{table}[bhtp]
	\centering
	\begin{tabular}{c c c c}
		\hline
		$bM$ & $r_{\rm ISCO}/M$ & ${\cal L}$ & ${\cal E}$\\
		\hline
		\multicolumn{4}{c}{\textbf{Prograde} (${\cal L}>0$)} \\
		\hline
		0.000 & 3.00000 & $+1.61185$ & 0.93060 \\
		0.001 & 2.99981 & $+1.61216$ & 0.93077 \\
		0.002 & 2.99925 & $+1.61254$ & 0.93096 \\
		0.003 & 2.99832 & $+1.61301$ & 0.93119 \\
		0.004 & 2.99702 & $+1.61357$ & 0.93144 \\
		0.005 & 2.99537 & $+1.61420$ & 0.93172 \\
		\hline
		\multicolumn{4}{c}{\textbf{Retrograde} (${\cal L}<0$)} \\
		\hline
		0.000 & 3.00000 & $-1.61185$ & 0.93060 \\
		0.001 & 2.99981 & $-1.61164$ & 0.93047 \\
		0.002 & 2.99925 & $-1.61150$ & 0.93036 \\
		0.003 & 2.99831 & $-1.61145$ & 0.93029 \\
		0.004 & 2.99701 & $-1.61149$ & 0.93024 \\
		0.005 & 2.99534 & $-1.61160$ & 0.93022 \\
		\hline
	\end{tabular}	\caption{ISCO data for $k=\tfrac12$ separated into prograde (${\cal L}>0$) and retrograde (${\cal L}<0$) branches.}
	\label{tab:isco-khalf}
\end{table}

\section{Null circular motion and photon orbits}
\label{sec:null-effective}

We now examine equatorial null geodesics (photon motion) using the same framework established in Sec.~\ref{sec:charged-test}. For a massless particle with four-momentum $p_\alpha$, stationarity and axisymmetry imply two conserved quantities,
\begin{equation}
E\equiv -p_t,\qquad L_z\equiv p_\phi,
\end{equation}
and the null condition reads $p_\alpha p^\alpha=0$. Restricting the analysis to the equatorial plane $\theta=\pi/2$ and using an affine parameter $\lambda$ (so $\dot{}\equiv d/d\lambda$), the null condition yields
\begin{equation}
g_{rr}\,\dot r^{\,2}+g^{tt}E^2+g^{\phi\phi}L_z^2=0.
\label{eq:nullHJform}
\end{equation}
This is the massless counterpart of Eq.~\eqref{eq:HJ-radial-compact}. It is useful to cast it in an ``energy minus potential'' form analogous to the timelike case. Defining the impact parameter
\begin{equation}
d_{\rm imp}\equiv \frac{L_z}{E},
\end{equation}
Eq.~\eqref{eq:nullHJform} becomes
\begin{equation}
g_{rr}\,\dot r^{\,2}
= -g^{tt}E^2\left[1+\frac{g^{\phi\phi}}{g^{tt}}\,d_{\rm imp}^{\,2}\right]
= \frac{E^2}{(-g_{tt})}\left[1+\frac{g_{tt}}{g_{\phi\phi}}\,d_{\rm imp}^{\,2}\right].
\label{eq:null_radial_imp}
\end{equation}
Equivalently, one may introduce the null effective potential at fixed angular momentum $L_z$,
\begin{equation}
\mathcal R_{\rm null}(r)\equiv \mathcal E_{\rm n}^{2}-U_{\rm null}(r;d_{\rm imp}),
\qquad
\mathcal E_{\rm n}\equiv \frac{E}{|E|}=1,
\qquad
U_{\rm null}(r;d_{\rm imp})\equiv - \frac{g_{tt}}{g_{\phi\phi}}\,d_{\rm imp}^{\,2},
\label{eq:null_Reff}
\end{equation}
so that turning points satisfy $\mathcal R_{\rm null}=0$.

Circular photon orbits at $r=r_\gamma$ are defined by the simultaneous conditions $\dot r=0$ (turning point) and the extremum of the potential. A convenient invariant way to impose these is to work with the squared impact parameter function,
\begin{equation}
d_{\rm imp}^{\,2}(r)=\frac{g_{\phi\phi}}{-g_{tt}},
\label{eq:bimp_def}
\end{equation}
together with the derivative condition
\begin{equation}
\frac{d}{dr}\!\left(d_{\rm imp}^{\,2}\right)\Big|_{r_\gamma}=0.
\label{eq:photon_condition}
\end{equation}
Equation~\eqref{eq:photon_condition} is the null analogue of the circular-orbit conditions $\mathcal R=0$ and $\mathcal R'=0$ used for timelike motion. 
In general, Eq.~\eqref{eq:photon_condition} only identifies \emph{extrema} of $d_{\rm imp}^2(r)$ (equivalently extrema of $U_{\rm null}$ at fixed $L_z$); whether the corresponding circular null orbit is stable or unstable requires an additional check of the second derivative (e.g.\ $(d_{\rm imp}^2)''(r_\gamma)$ or $U_{\rm null}''(r_\gamma)$). 
In particular, in the Melvin--Schwarzschild limit ($k\to 1$) stable circular null geodesics are known to exist for sufficiently strong magnetization (see, e.g., Refs.~\cite{Stuchlik:1999mro,Lim:2015oha}); throughout this work we do not attempt a global classification of null-orbit stability for general $k$, and we focus instead on the ``photon-ring--like'' branch continuously connected to the unmagnetized Zipoy--Voorhees photon ring.

Using the equatorial metric functions given in Eq.~\eqref{eq:eq-metric-Aphi}, we obtain the explicit expression for the squared impact parameter
\begin{equation}
d_{\rm imp}^{\,2}(r)=\frac{g_{\phi\phi}}{-g_{tt}}
=\frac{r^{2}\,f^{\,1-2k}}{\Lambda^{4}}.
\label{eq:bimp_correct}
\end{equation}
The photon ring radius $r_\gamma$ is then determined by numerically solving Eq.~\eqref{eq:photon_condition} for the extrema of Eq.~\eqref{eq:bimp_correct}. The unmagnetized limit reproduces the standard Zipoy--Voorhees baselines; in particular, we recover $r_\gamma=3M$ for the Schwarzschild case ($k=1$). The results for weak magnetic fields are displayed in Table~\ref{tab:photon_radii}. Notably, increasing the magnetic parameter $bM$ shifts the equatorial photon ring slightly outward. This behavior stands in contrast to the timelike ISCO radii in Tables~\ref{tab:isco-k2}--\ref{tab:isco-khalf}, which shift inward as $b$ increases. This divergence in behavior is physically expected: magnetization modifies the null potential ratio $(-g_{tt})/g_{\phi\phi}$ directly through the conformal factor $\Lambda$, whereas for massive charged particles, the stability boundary is additionally sensitive to the Lorentz coupling encoded in the gauge-invariant angular momentum $\tilde{\mathcal L}=\mathcal L-\eta A_\phi$, which dominates the dynamics near the ISCO.

\begin{table}[hbtp]
	\centering
	\begin{tabular}{c c c c c c c}
		\hline\hline
		$k$ \textbackslash $b M$ & 0.0 & 0.001 & 0.002 & 0.003 & 0.004 & 0.005 \\[0.5ex]
		\hline
		0.8 & 2.600000 & 2.600004 & 2.600016 & 2.600036 & 2.600065 & 2.600101 \\
		1.0 & 3.000000 & 3.000009 & 3.000036 & 3.000081 & 3.000144 & 3.000225 \\
		2.0 & 5.000000 & 5.000083 & 5.000333 & 5.000750 & 5.001333 & 5.002083 \\
		\hline
	\end{tabular}
	\caption{Equatorial photon ring radius $r_\gamma/M$ in the Melvin--Zipoy--Voorhees spacetime as a
		function of the deformation parameter $k$ and the magnetic field strength $b M$. Radii are
		determined by numerically solving the extremum condition~\eqref{eq:photon_condition} for
		$d_{\rm imp}^2(r)=g_{\phi\phi}/(-g_{tt})$ in Eq.~\eqref{eq:bimp_correct}.}
	\label{tab:photon_radii}
\end{table}

\section{Conclusion}
\label{sec:conclusion}

We have constructed a magnetized generalization of the Zipoy-Voorhees geometry by applying the magnetic Harrison transformation to a static, axially symmetric seed endowed with quadrupolar deformation. The resulting Melvin-Zipoy-Voorhees spacetime is an exact solution of the Einstein-Maxwell equations that smoothly interpolates between two fundamental limits: it recovers the unmagnetized Zipoy-Voorhees metric in the absence of a magnetic field and approaches the Melvin magnetic universe in the weak-gravity, quasi-spherical regime. The solution is characterized by a purely azimuthal vector potential, yielding an electromagnetic field that is strictly magnetic for static observers with a vanishing electromagnetic pseudo-invariant, a property we verified explicitly using an orthonormal tetrad adapted to the static congruence.

From a curvature perspective, the spacetime is generically of Petrov type I, specializing to type D only in the Schwarzschild limit. Our analysis of the Kretschmann scalar on the equatorial plane reveals a coherent interplay between the intrinsic deformation and the external magnetic field: while increasing the deformation enhances the curvature near the source, magnetization systematically suppresses the equatorial curvature amplitude across a broad radial domain. These features are consistent with the redistribution of local energy density between the gravitational and electromagnetic sectors characteristic of magnetized Weyl geometries.

The dynamics of test particles on the equatorial plane are governed by two complementary magnetic mechanisms. First, the vector potential induces a Lorentz shift in the mechanical angular momentum of charged particles, which is linear in both the charge-to-mass ratio and the magnetic field strength; this interaction effectively modifies the centrifugal barrier, either weakening or strengthening it depending on the alignment of the charge and the external field. Second, the Melvin magnetization factor rescales the temporal and azimuthal metric components, modifying the effective potential quadratically in the magnetic field strength. Together, these effects reshape the effective potential well, thereby controlling the existence and stability of circular orbits. For massless particles, the condition for circular photon motion is determined by the extrema of the squared impact parameter. We found that the deformation parameter establishes the baseline radius of the photon ring in the unmagnetized limit, while the external magnetic field induces a smooth, systematic shift in this radius across the parameter space.

The Melvin-Zipoy-Voorhees solution thus serves as a compact analytic laboratory for disentangling how external magnetic fields and intrinsic quadrupolar deformations jointly sculpt the geodesic structure and physical observables around compact objects. Possible works for future investigations include a comprehensive mapping of innermost stable circular orbits (ISCOs) for charged particles, an investigation of off-equatorial dynamics and orbital resonances, and the computation of ray-traced images and shadow morphologies in the presence of the external field. Further extensions could explore the quasi-normal mode spectra and stability against scalar and vector perturbations, as well as the integration of simple plasma models to predict polarimetric and high-energy signatures. These studies will help refine our understanding of the roles played by external magnetization and multipolar structure in realistic astrophysical environments, clarifying which observational signatures are robust across families of deformed, magnetized spacetimes.

	\section*{Acknowledgements}
	
	The author is supported by LPPM-UNPAR through Publikasi Internasional Bereputasi scheme.

\end{document}